\documentclass[12pt, draftclsnofoot, onecolumn]{IEEEtran}
\usepackage{amssymb}
\usepackage{amsmath}
\usepackage{amsfonts}
\usepackage{graphicx}
\usepackage{algorithm}
\usepackage{algorithmic}
\usepackage{epstopdf}
\usepackage{cite}
\usepackage{color}
\usepackage{multirow}
\usepackage{subfigure}
\usepackage{bm}
\usepackage{hyperref}

\begin{document}
\title{When ICN Meets C-RAN  for HetNets: An SDN Approach}
\author{Chenchen Yang, {\em Student Member, IEEE}, Zhiyong Chen, {\em Member, IEEE}, Bin Xia, {\em Senior Member, IEEE}, Jiangzhou Wang, {\em Senior Member, IEEE}\\
\thanks{This work has been accepted by IEEE Communications Magazine. C. Yang, Z. Chen, and B. Xia (corresponding author) are with the Department of Electronic Engineering, Shanghai Jiao Tong University, Shanghai,  P. R. China. }
\thanks{J. Wang is with School of Engineering and Digital Arts, University of Kent, United Kingdom.}
}

\maketitle

\begin{abstract}
With the ever-growing of mobile Internet and the explosion of its applications, users are experiencing  abundant services from different content providers via different network service providers in the heterogeneous network (HetNet). Information-centric network (ICN) advocates to get rid of the current host-centric network protocol for the fact that the information dissemination rather than the end-to-end communication contributes to the majority of today's network traffic. Furthermore, it is better for network entities to converge as a whole in order to take advantage of the open, scalability, and smart traffic transmissions. Cloud radio access network (C-RAN) is one of the emerging architecture evolutions in the wireless side for easier infrastructure deployment and network management. Therefore, when ICN meets C-RAN in the HetNet, it is worthwhile and consequential to integrate ICN protocol with the C-RAN architecture to achieve more efficient communication and information management. Moreover, software defined networking (SDN) has been recognized as another promising architecture evolution to achieve the flexibility and reconfigurability of dense HetNets, whose inherent advantage yet lies in the global uniform control of the wired network. Thus any of ICN, C-RAN and SDN is the complementary to the others.
In this paper, we contribute  to novelly proposing and elaborating the integration of the ICN, C-RAN and SDN for the HetNet to achieve win-win situation. The vision of the proposed  system is demonstrated, followed by the advantages and challenges. We further present the hybrid system with a large-scale wireless heterogeneous campus network.
\end{abstract}
\section{BACKGROUND AND MOTIVATIONS}
To meet with the thousand-fold growth of mobile data traffic, a promising way is embedding mass of small cells into the cellular network, i.e. the heterogeneous network (HetNet) \cite{5Gmultitier}. Heterogeneity will be an essential feature of the future radio access network (RAN) consisting of different kinds of devices.
For the highly flexible interfaces among different devices, software defined networking (SDN) has been recognized as a basic enabler to achieve the flexibility and reconfigurability of dense HetNets \cite{SDN1}. Furthermore, under the pressure of inter-/intra-cell coordination and tremendous deployment cost in the large-scale HetNet, Cloud RAN (C-RAN) emerges \cite{CRAN3}. C-RAN is widely regarded as a green radio architecture to cope with the increasing network data traffic without incurring significantly additional capital and operational expenses. In addition to the evolution of the network architecture, the human behavior and the service property have been widely concerned to unleash the ultimate potential of networks, e.g., the cache-enabled Information-centric network (ICN) \cite{femto2}. The state of the art will be elaborated in the following in the perspective of network architecture and  network protocol, respectively.
\subsection{From the Perspective of Network Architecture}
\textbf{The HetNet is the expansion of the existing network in terms of  the size and the type:}
Compared with the the fourth Generation/long term evolution (4G/LTE) network, the fifth Generation (5G) network is expected to attain a 1000-fold capacity increase measured in bit/s/Hz/m$^2$, 5 times reduced latency and 10 times longer battery life \cite{metics}. Motivated by the ever-increasing pressure to enhance the network capacity,  cellular networks are overlaid by a wide variety of proliferative infrastructure layers including outdoor pico/indoor femto  base stations (BS), relays,  distributed antennas\cite{wang1}, WiFi and device to device (D2D) nodes \cite{5G4}. Heterogeneity will be a key attribute of the future networks.

HetNets bring the cell site closer to the users and  shorten the radio transmission distance, yielding significant cost reduction and capacity  enhancement.
Furthermore, the small cells are deployed as the complementary of the macro cellular networks for the seamless coverage and intentionally offload the traffic of the macro cell. Besides, advanced technologies such as high-order spatial multiplexing multiple-input multiple-output (MIMO) come along with higher spectral efficiency (SE) in the HetNet. However, there are still problems:
\begin{itemize}
  \item The resource management and load balancing are two important issues for the successful deployment of HetNets. The inter-/intra-cell coordination and convergence are non-trivial operations in this distributed and large-scale environment.
  \item HetNets enable more flexible and economical deployments of new infrastructure instead of tower-mounted macro systems, but it is not sufficient to achieve the flexibility and reconfigurability of the networks due to the nonuniform and curable device interfaces.
  \item The increasing demands on the backhaul network  needs to be evaluated when deploying more small cell sites.
\end{itemize}
The trivial expansion of the size and type of the exiting network can temporarily relieve the pressure, but disruptive evolutions of the network architecture are urgent for the further capacity extension.

\textbf{C-RAN is the architecture evolution in the wireless network side:}
Unlike the existing HetNet, where  each transport node (e.g., Macro/Micro BS, relay, etc.)
posses computing resources for baseband processing in its local cell site, C-RAN aims to separate  radio access units (RAUs) and baseband processing units (BBUs). C-RAN consists of three fundamental components: (i) the distributed RAUs equipped with remote radio heads (RRHs) at the cell site; (ii) the centralized BBU pool performed in a data center cloud; (iii) the intermediate entity unit providing the high-bandwidth low-latency links to connect RRHs and BBU pool. C-RAN gathers a large number of BBUs into a logically centralized BBU pool. The centralized and coordinated baseband processing/scheduling in the BBUs allows for soft and dynamic cell reconfiguration. The light-weight RRHs  is implicitly decoupled from the BBU pool in the C-RAN,  yielding the easier deployment of different types of cells in the HetNet, and reducing the energy consumption of each site.  RRHs cooperate flexibly and seamlessly in the HetNet so the SE and the capacity can be improved significantly.

However, challenges come along with the advantages: full-scale coordination leads to high computational overhead in the BBU pool especially for the large-scale network; real-time virtualization and high-bandwidth links are urgent to achieve the reliable connection and mapping between the BBU and the RRH. C-RAN only focuses on radio air interface and cannot solve the problems emerging in the core network (CN) or other upper layers. But C-RAN can be regarded as a specific sub-unit of the SDN which will be elaborated in the following.

\textbf{SDN is the architecture evolution  in both the wired and wireless sides:}
Tremendous number of infrastructure equipments have been patched to the network nowadays. Traditionally, each device is  an ossified unity of the hardware and software developed by the unique manufacturers. Every new feature or capability expansion requires professionally manual update and reconfiguration of the software stack with proprietary languages. SDN decouples the control plane from the data plane via \emph{controllers} and allows software to be designed independently from the hardware,  simplifying the network access, design and operation \cite{ref2}. It fosters the network virtualization and makes it  possible to evolve the consolidation of different types of network equipment, which reduces the  network complexity for sophisticated heterogeneous systems.

SDN provides an open interface with centralized \emph{controllers} remotely controlling the data forwarding tables in the network entities such as switches, routers, and access points. The network entities  are virtualized and solely take forwarding strategies defined by the external programmable \emph{controllers}.  Logically centralized \emph{controllers} possess the flexibility to dynamically and automatically redirect or optimize the traffic load and resource management with the global view of the system.

However, SDN is originally designed for wired networks. There are inherent weaknesses to achieve the SDN ideas in the wireless networks due to vary challenges. Software defined wireless networks (SDWN) needs to define slices which requires to isolate wireless channels so as to provide non-interfering networks to different coordinators. Handoff situations should be considered for the wireless HetNet with smaller cells and richer access technologies. The status and locations of the whole network entities should be reported to SDN in real time only based on which  \emph{controllers}  take decisions efficiently. These challenges are no-trivial operations for the wireless medium. There is a long way for the wired and wireless sides to be seamlessly integrated as a whole in the SDN system.
\subsection{From the Perspective of Network Protocol}
\textbf{ICN provides the new basis on  how information can be labeled and distributed across networks:}
Despite of the tremendously amount of data traffic in the network, only a few contents are frequently accessed by user ends \cite{Zipf}. A small portion of the popular contents contribute to the majority of the traffic during a period of time \cite{kongtao}. Furthermore, Today's network is increasingly occupied by information dissemination, rather than by the end-to-end communications. Thus caching the most popular contents in the RAN or the evolved packet core (EPC) can reduce the redundant access and the duplicated transmission \cite{5G1}. The pivotal role the caching technology can play for the 5G wireless network is elaborated in \cite{5G1} and \cite{5G3}. ICN emerges as a promising candidate for full use of in-network caching and multicast mechanisms \cite{ICN1}.

Different in the exiting host-centric networks where users have to transmit/receive information to/from a particular computing entity (host or server), they are increasingly interested in transmitting/receiving information wherever they may be located. The ICN decouples information from its location and sources by defining the named data objects (NDO).  Widespread caching and broadcasting allow users to get the information from the optimal node based on the name and/or the location of the information. Providers and requesters are no longer connected in the traditional pair-wise and time-synchronized mode, they are decoupled in terms of time and space. There is no need for them to know each other's location or to be online at the same time.
\subsection{Contribution}
As mentioned above, the widespread in-networking caching of ICN brings great opportunities of cooperative communication among entities in the high-density wireless HetNet, while C-RAN is emerging for easier infrastructure deployment and system management. Then integrating ICN protocol into the C-RAN architecture can achieve better communication with efficient distribution of information via ubiquitous cache-enabled devices. At the same time, the burden of the upper layer (e.g., the CN) will be significantly reduced when user requests can be responded immediately by the entity (e.g., the BBU) who has cached the copy of the requested information.  However, C-RAN solely focus on the radio air interface regardless of the states of upper layers. Coincidentally,  SDN is born with the  original purpose of dealing with wired networks but it will meet more difficulties to introduce concepts of SDN to wireless networks. Moreover, the ICN protocol is available to the SDN as well. Therefore, appropriate combination of ICN, CRAN and SDN is a promising way to achieve mutual benefit and complementary.

We organize the reminder of this paper as follows: In Section \uppercase\expandafter{\romannumeral2}, we describe our vision of the coexisting system of the SDN, ICN and H-CRAN. The advantages and main challenges are elaborated. In Section \uppercase\expandafter{\romannumeral3}, a large-scale exemplary campus network is presented. \uppercase\expandafter{\romannumeral4} gives a brief summary of this article.
\section{The Vision of the Future Network System}
The Information-centric concept is better suited to today's use which mainly consists of information distribution rather than host-centric communications. For the long tail style of the user request statistical property, in-network caching and multiparty communication can be fully leveraged via replication and cooperative models.  We propose to achieve the information-centric approache on the SDN architecture to unleash the ultimate potential of the network, and C-RAN  undertakes  the role and task of SDN in the wireless side. Few of existing studies has considered the coexisting network and we will give an integrate sight of them.
\subsection{Information-centric Software Defined Networking}
\begin{figure}[t]
\centering
\includegraphics[width=5.5in]{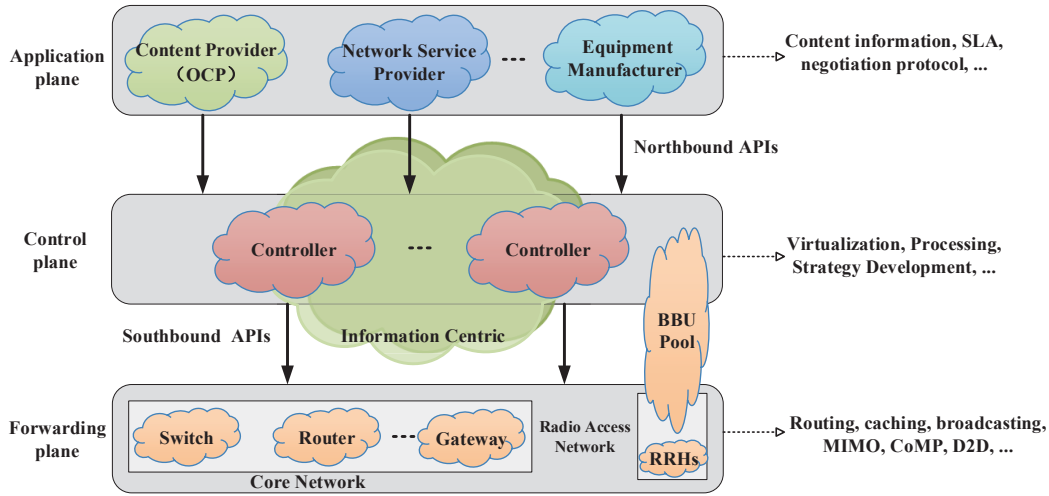}
\caption{The information-centric SDN with C-RAN.}
\label{structure}
\end{figure}

A high-level view of information-centric SDN architecture is shown in Fig. \ref{structure}, where the application plane, control plane and forwarding plane are included. The forwarding and control plane components can evolve independently via defining standard application programming interfaces (APIs) between them.

The application plane consists of the application and service provided by the following three entities: content provider (CP) which contains the traditional CP and the emerging over-the-top (OTT) content provider (OCP) such as Google, Amazon and Netflix; network service provider (NSP) and the equipment manufacturer (EM). The control plane consists of a set of distributed but logical centralized \emph{controllers}.  A \emph{controller} can control quantity of network entities and a network entity  can be controlled by different but logically centralized \emph{controllers}. However, there is no need to provide a strictly consistent centralized view to each \emph{controller}, which will cause processing overload and additional cost when the network expands.  A control functionality even can be completed by different \emph{controllers}. Therefore, selecting an appropriate consistency level (e.g., the weakest possible consistency level) of \emph{controllers} is an important design consideration in SDN to preserve scalability \cite{scrability}.
The CN and the RAN are included in the forwarding plane. The infrastructure of the CN can be virtualized and controlled by the control plane. The centralized BBU pool and distributed RRHs of C-RAN are deployed  to achieve efficient collaboration among different cells in HetNets. Therefore, independent networks can be reconfigured flexibly and automatically on the same physical infrastructure under the `soft' strategy change of \emph{controllers}.
The role that each plane plays in the network is :
\begin{itemize}
\item Application plane: The CP (e.g., OCP)  distributes the NDO in its server to user ends with the help of NSPs. The objective/content information is decoupled from its location and sources by naming approach of ICN, thus it is more suitable for the CP to understand and forecast user behaviors (e.g., the request frequency and the content population distribution). Given the service-level agreement (SLA) provided by different NSPs, the CP choose the optimal NSPs via negotiation to distribute the objective. Decoupling the software and the hardware via SDN may help  CPs and NSPs get rid of the shackles of the EM. It is not what the EM wants and negotiation protocol also should be developed  among them. The application requirements are negotiated based on the protocol running in the underlying control plane through open APIs.
\item Control plane: The control plane running in a network operating system (NOS) is the core part of the  architecture, and the logically centralized \emph{controllers} complete the infrastructure virtualization, programming abstractions, even the content naming, addressing and matching procedures for the ICN.  The \emph{controller} exploits complete knowledge of the system and get consolidated control functions to facilitate the networks reconfiguration and management via the NOS. For example, wired backhual and wireless bandwidth owned by the NSPs can be dynamically allocated to the CPs and users respectively based on the negotiation protocol running in the \emph{controller}.
\item Forwarding plane: Since the control function has been extracted and integrated into the control plane, the forwarding plane thus consists of simplified and virtualized network devices that solely provide information switching and forwarding. However, a typical HetNet may have tens of thousands of devices and the sheer number of control events generated at that scale is enough to overload any \emph{controller}. pushing all the control functionality to centralized \emph{controllers} makes them the potential bottleneck for the network operation when the HetNet scales.
\end{itemize}

To reduce the burden of the \emph{controllers} as well as to get rid of the aforementioned no-trivial challenges of SDWN, we propose to deploy the wireless side of SDN in the concept of C-RAN  where the BBU pool has both control and data forwarding functions. The logically centralized BBU pool has the network-wide view of the RAN and the CN, yielding the seamless integration of the wired and wireless sides of SDN. Collaboratively controlled by the \emph{controllers} and the BBU pool, the NDO of ICN can be flexibly and optimally distributed and stored in diverse devices of the CN and the RAN via caching and broadcasting.
\subsection{The deployment of C-RAN in the information-centric SDN}
Embedding C-RAN into the information-centric SDN is rather a promising  way to reduce the burden of  \emph{controllers}  and to integrate the wired and wireless sides of SDN seamlessly. It is more efficient and easier to set up better broadcasting/multicasting and in-network caching mechanisms (e.g., information placement and replacement strategy) which are cornerstones of ICN.

\begin{figure}[t]
\centering
\includegraphics[width=0.8\linewidth]{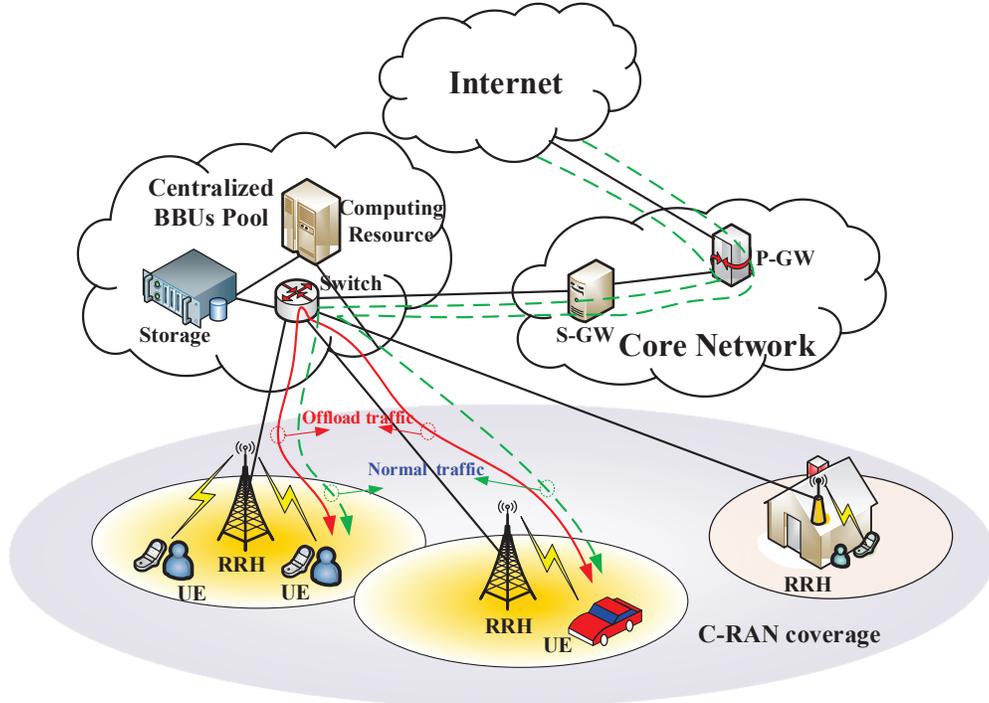}
\caption{An overview of offload traffic.}
\label{offload}
\end{figure}

\begin{figure}[t]
\centering
\includegraphics[width=5.0in]{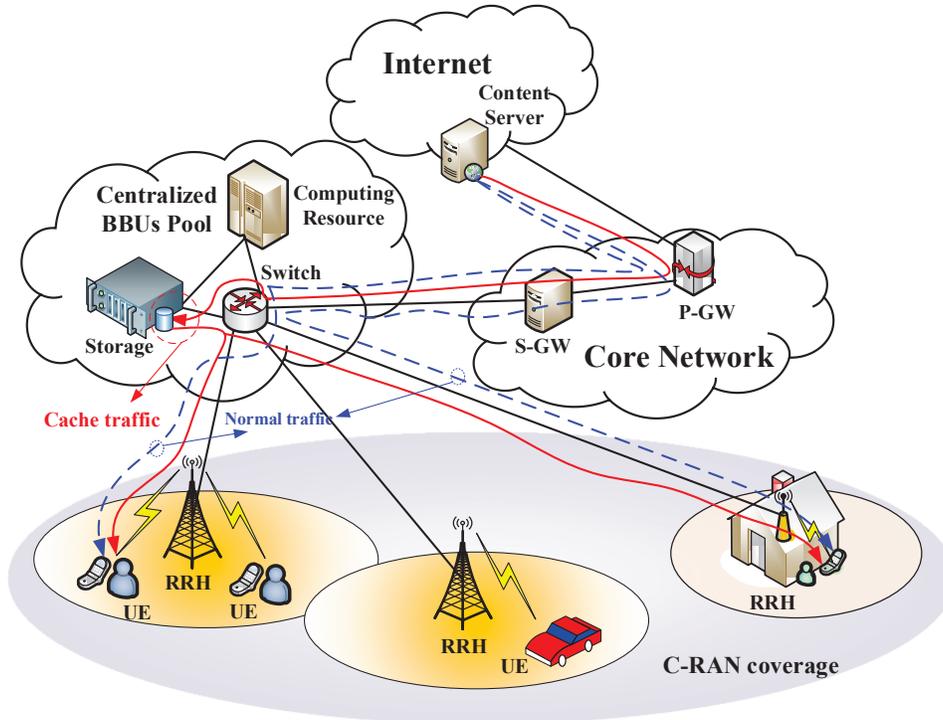}
\caption{An overview of cache traffic.}
\label{structure2}
\end{figure}

Without doubt, the explosive traffic demand increases the delivery cost of mobile Internet content on the operator. With the help of `cloud'  resources, C-RAN involves all the computing, storage, and content-aware elements required to functionally and efficiently deliver the content, which enables new smart traffics of ICN such as offload traffic and cache traffic as depicted in Fig. \ref{offload} and Fig. \ref{structure2} to meet the challenge.
\begin{itemize}
  \item \emph{Offload traffic}: a promising solution that offloads the data traffic bypassing the CN of the operator and Internet. As illustrated in Fig. \ref{offload}, when the user equipment (UE) communicates, the normal traffic is required to traverse the RAN, the CN and Internet before reaching other UE. For the offload traffic, the centralized BBU pool enables a Clould-identification (ID)-enabled UE, connected via RRH,  to directly connect to other UE configured with the identical Cloud-ID without passing the CN and Internet. Note that such offload traffic is established at the RAN, which is different from the traditional offload approach enabling data traffic offload via Wireless local area network (WLAN). Through enabling the data traffic closer to the edge of network (i.e., RAN), the operator can relieve the pressure on the CN elements such as the Serving gateways (S-GW) and the Packet Data gateways (P-GW), yielding the cost saving of gateway upgrades. Moreover, the offload traffic allows the direct communications with UEs within one C-RAN network, reducing the content latency and efficiently improving the quality of service (QoS) and user experience. Besides, the two UEs even can communicate directly without traversing the BBU pool in the mature ICN which will bring more benefit.
  \item \emph{Cache traffic}: Fig. \ref{structure2} demonstrates a positive solution that selectively pushes and stores the mobile contents at the centralized BBU pool, supporting the enhancement to directly access the contents without from the content servers at Internet via the CN. Meanwhile, following the uncannily accurate Moore's law, the storage and processing capability of the intelligent devices becomes more and more stronger. When the network is at off-peak traffic load, e.g., in the night time, the most frequently accessed content can be broadcasted and then cached at the BBU pool. The user can obtain the requested content immediately from the BBU pool when the cache hit event occurs. As a result, the deployment of the C-RAN infrastructure in the information-centric SDN  leverages the storage in the network edge for caching to provide more and richer services, whereby the cache is independent of the application and can be applied to various sources, including user-generated content, which is an attractive feature of ICN.
\end{itemize}

In addition to caching in the BBU pool, the  popular contents can also be pushed and cached in the RAN or the EPC when the network is idle. It offloads the corresponding redundant traffic and the duplicated access in the CN and the RAN, which can reduce the pressure of the backhaul link and the processing overhead in the BBU pool and \emph{controllers}. For example, NSPs can proactively pushes and caches popular contents published by  CPs to the cache-enabled intermediate proxy devices or part of the UEs in advance, then the UE can get the content from the optimal node of ICN via D2D or the proxy-UE link when time is available.
\subsection{Two Dimensions and Main Challenges}
\begin{figure}[t]
\centering
\includegraphics[width=6.0in]{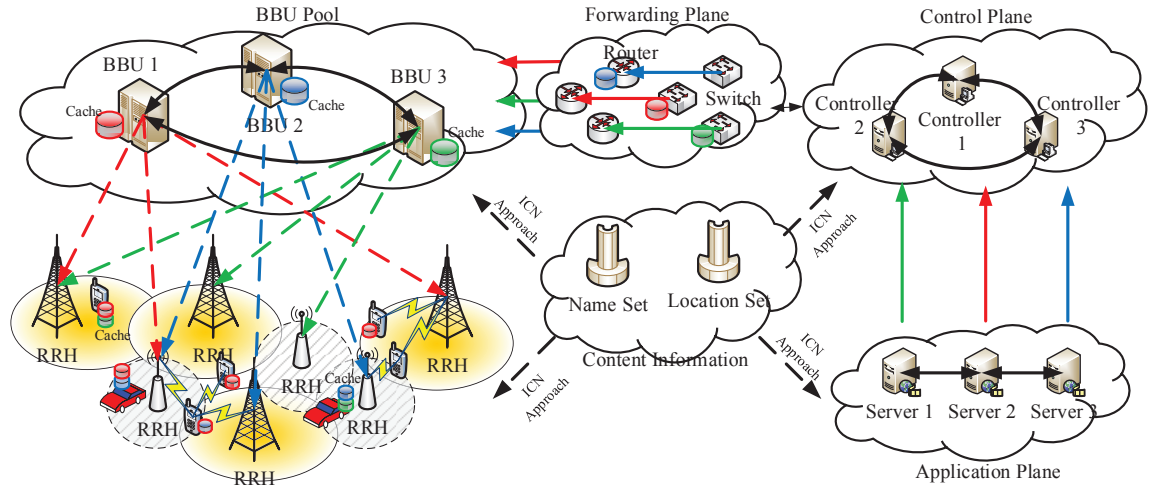}
\caption{The coexisting system of the SDN, ICN and H-CRAN.}
\label{structure3}
\end{figure}

Thus, in order to get rid of the obstacles in the development of SDWN and make use of the efficient transmission of ICN, we propose a new ICN, SDN and H-CRAN integrated system as described in Fig. \ref{structure3} which encompasses the complete system platform. The novelty of the proposed concept can be elaborated from the following two dimensions.

\begin{itemize}
\item The architecture dimension: Infrastructures are deployed with the devices whose software are decoupled from the hardware and centralized to the control entity (i.e., BBU pool and \emph{controllers}). Different CPs provide abundant services to subscribers with the help of different NSPs based on the negotiation protocol which is well known to the control plane. \emph{Controllers} are the intermediate entities between the applications plane and the forwarding plane. The necessary state message of the control plane and the forwarding plane should be reported to \emph{controllers} via APIs. Appropriate forwarding strategies are developed by the control plane with the global view of the network (BBU pool can cooperate with \emph{controllers}). The control plane distributes the incoming traffic provided by the servers of the application plane to the forwarding plane. Entities such as routers solely take the strategy developed by the \emph{controllers} and forward traffic to C-RAN. Non-uniformly distributed traffic and requests can be balanced in the BBU pool and then transmitted to UEs at last.
\item The information dimension: The main abstraction of information in ICN is the NDO for identifying the information independent of its location or publisher.  The role the NDO plays in the ICN is as important as that the internet protocol (IP) plays in the current host-centric Internet. When the information labeled with unique NDOs has been published  by the sever in the application plane or the UE in the RAN, its transcript can be held by ubiquitous cache-enabled nodes (e.g., nodes in EPC or RAN) afterwards. For example, the transcript can be cached in BBU pool, proxies or UEs based on the caching strategy and the content access protocol developed by the BBU pool, yielding offloading the processing burden of \emph{controllers} and the traffic of the CN. Obviously, the design of the information dimension can affect the strategy and procedure in the architecture dimension.
\end{itemize}

However, there are still some open problems. Main challenges and future research directions can be summarized as follows:
\begin{itemize}
\item \textbf{Processing overload:} The huge performance gain of the proposed system mainly comes from the centralized and coordinated signal processing at the cloud BBU pool and \emph{controllers}. However, full-scale coordination in large-scale HetNets requires the processing of very large network information such as the channel matrices. Processing overload could be a major problem for such centralized environment as the network size grows even though the BBU pool and \emph{controllers} are able to cooperate with each other. One way to tackle this problem is to decrease the redundant and duplicated  flow before they enter the control plan with appropriate strategy (e.g., optimal caching in ICN). Or as what \cite{scrability} proposes, to proactively install rules on the devices to  eliminate some control requests, while yielding some loss of control precision and reactivity.
\item \textbf{Backhual and Fronthaul:} The common assumption that the  data can be routed to the RANs without  backhaul and Fronthaul limitation is not valid for the future high-density HetNet,  where a large number of nodes need to access information. High-capacity wired backhual and Fronthaul are needed for the connections of the application plane, the control plane and the forwarding plane (e.g., the \emph{controllers} and CPs, the BBUs and the RRHs). Few existing  studies jointly consider the wired backhaul, fronthaul and the radio resource management in HetNets  when optimizing the system performance.
\item \textbf{Universal Caching:} The universal in-network caching is a salient feature of the proposed system, for which the caching strategy and content replacement  are important issues. Caching strategy decides what, where, when and how the information should be cached to achieve the optimal system performance. Furthermore, the cached information should keep consistent to those in the server and the publisher so the invalid information should be replaced with appropriate mechanism in real time. User behaviors (e.g., the access frequency follows the long tail distribution) are not yet fully investigated especially in the dense mobile HetNet, but it can significantly influence the  performance of the caching strategy.
\item \textbf{Access protocol and data routing:} The proposed system should apply to any  access protocol in the sophisticated HetNet, not just a specific protocol (e.g., HTTP, LTE). It should thus provide a uniform content distribution paradigm underlying all access protocols. Moreover, flexible and convenient  information-aware mechanisms should be developed for the data routing based on the  location-independent name of the information. Then the subscriber can be responded by the optimal node who has cached the information rather than only by the original publisher.
\end{itemize}
There are some other challenges such as mobility and security management. With the RRHs deployed more and more concentrated in the future HetNet, it becomes more frequent for a user to handoff between different RANs.  On the other hand, similar to traditional security techniques such as transport layer security (TLS), equivalent security measures should be developed for naming objects, caching and communication in the proposed ICN. The key question is that the requesters can get content from any ubiquitously cache-enabled entity who have different capabilities other than the relatively uniform host servers, so the security measure should be based on the content itself (e.g., naming) rather than the communication channel or the path.
\section{IMPLEMENTATION of an Exemplary Network}
\begin{figure*}[t]
\centering
\includegraphics[width=5.5in]{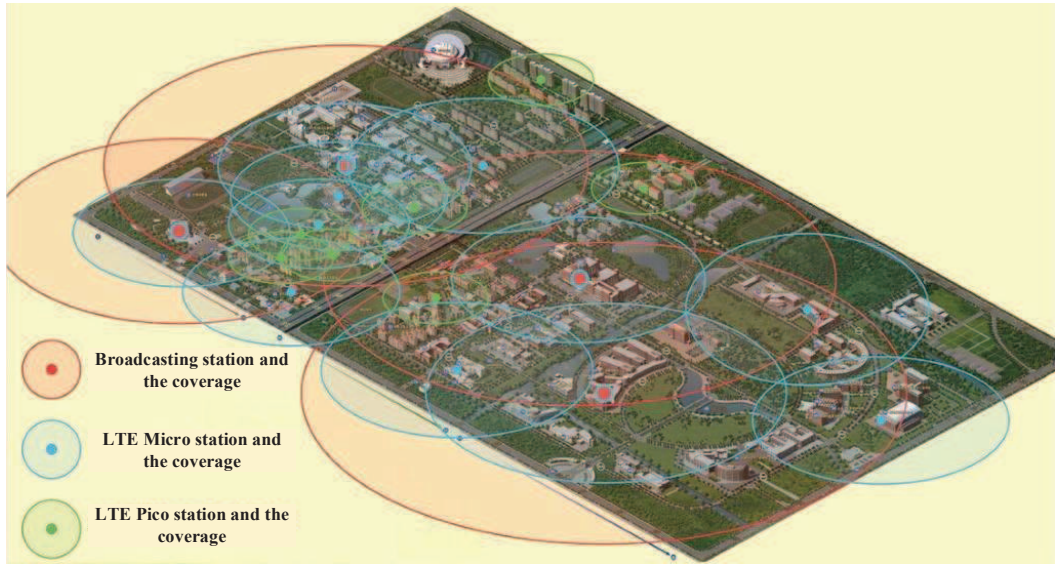}
\caption{A large-scale wireless heterogeneous campus network.}
\label{model_fig2}
\end{figure*}
We have established a large-scale wireless innovation campus network with 3 square kilometer coverage in Shanghai Jiao Tong University. The heterogeneous campus network consisting of a digital broadcasting system, a LTE cellular system, and a WiFi system is demonstrated in Fig. \ref{model_fig2}. The important features of the wireless heterogeneous campus network  are summarize as follows:
\begin{itemize}
    \item There are more than 80 LET Micro- and Pico-  stations with  blanket coverage of the campus. More than 2500 WiFi access points are patched across the entire campus. Additional 3 stations constitute a single-frequency digital broadcasting network.
    \item All cellular RF transceivers are connected to a computing center through over 66 kilometers of fiber, yielding a truly C-RAN.
\end{itemize}

Specially for the cooperative caching communication of SDN, each of the Pico stations and parts of UEs are cache-enabled.  Based on the strategy developed by the \emph{controllers} with the global view of the network, popular contents can be broadcasted to and then cached at the cache-enabled nodes. The cached content can be reused for the frequent access. Besides the cellular communication, there exists the device-to-device (D2D) link (i.e., from the cache-enabled users to the requesting users) for the content sharing, yielding a three-tier HetNet (i.e., Micro BSs-users, Pico BSs-users, D2D transmitters-users).  The UE in the overlapping coverage associates to the optimal node according to the instruction of the \emph{controllers}, and the UE can obtain the requested content immediately from its local caching disk if the content has been cached.

\begin{figure}[t]
\centering
\includegraphics[width=5in]{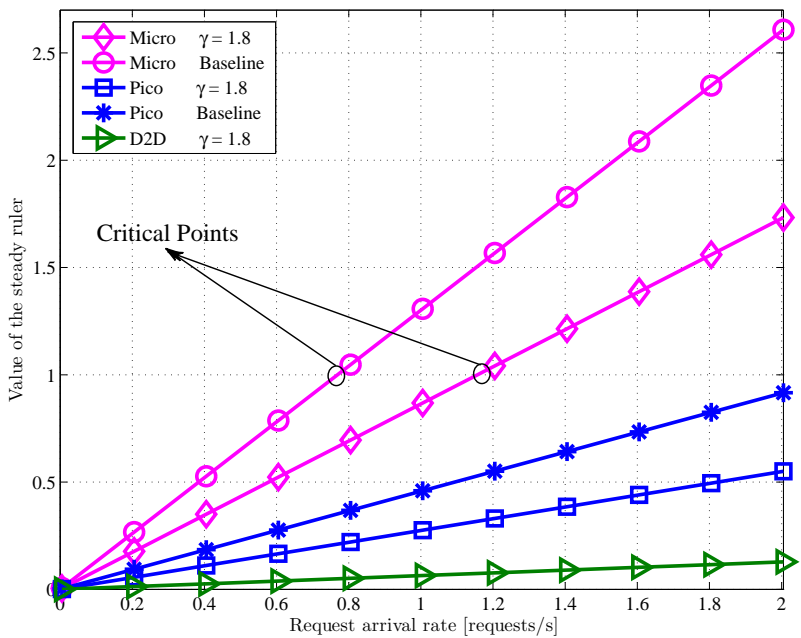}
\caption{The throughput gain for the cache-enabled network compared with that of the baseline.}
\label{steady}
\end{figure}

We verify the throughput gain of the coexisting system in \cite{twcmy}, compared with the baseline where there is not in-networking cache. The content access is triggered according to the well known Zipf distribution with parameter $\gamma$. Larger $\gamma$ implies that fewer contents account for the majority of the request.  In Fig. \ref{steady} we circle out the critical point deciding the maximum throughput of the network based on whether the steady ruler is larger than 1. We observe that when $\gamma=1.8$ the throughput gain is 53.9\% compared with that of the baseline. Moreover, the Pico and the D2D tiers are far from the fully-loaded state when the Micro tier comes to the critical steady state. Therefore, more appealing performance improvement can be further reaped with appropriate resource scheduling and load balancing mechanisms.

\section{Conclusions}
In this paper, to cope with the rapid  increase of network traffic and the change of the communication mode in the HetNet, we have presented a coexisting system  where the novel ICN, C-RAN and SDN are integrated to reap mutual benefit and complementary. Firstly, advantages and drawbacks of each kind of novel network are summarized respectively, based on which we have proposed an information-centric SDN architecture consisting of three planes. Network entities in each  plane have been explained and their corresponding roles have been clarified. Then we have elaborated the benefit that the C-RAN can bring to the information-centric system via examples of offload traffic and  cache traffic. Furthermore, the vision of the network platform  for the proposed system is described from architecture evolution and information dissemination points of view, at the same time challenges are enumerated. Finally, a demo large-scale wireless heterogeneous campus network is presented.
\section{Acknowledgment}
This work is supported in part by the National Natural Science Foundation of China (Grant No. 61301115) and the National High Technology Research and Development Program of China under 863 5G Grant 2014AA01A702.

\bibliographystyle{IEEEtran}
\bibliography{paper}

\begin{thebibliography}{10}
\providecommand{\url}[1]{#1}
\csname url@samestyle\endcsname
\providecommand{\newblock}{\relax}
\providecommand{\bibinfo}[2]{#2}
\providecommand{\BIBentrySTDinterwordspacing}{\spaceskip=0pt\relax}
\providecommand{\BIBentryALTinterwordstretchfactor}{4}
\providecommand{\BIBentryALTinterwordspacing}{\spaceskip=\fontdimen2\font plus
\BIBentryALTinterwordstretchfactor\fontdimen3\font minus
  \fontdimen4\font\relax}
\providecommand{\BIBforeignlanguage}[2]{{%
\expandafter\ifx\csname l@#1\endcsname\relax
\typeout{** WARNING: IEEEtran.bst: No hyphenation pattern has been}%
\typeout{** loaded for the language `#1'. Using the pattern for}%
\typeout{** the default language instead.}%
\else
\language=\csname l@#1\endcsname
\fi
#2}}
\providecommand{\BIBdecl}{\relax}
\BIBdecl

\bibitem{5Gmultitier}
E.~Hossain, M.~Rasti, H.~Tabassum, and A.~Abdelnasser, ``{Evolution toward 5G
  multi-tier cellular wireless networks: An interference management
  perspective},'' \emph{IEEE Trans. Wireless Commun.}, vol.~21, no.~3, pp.
  118--127, Jun. 2014.

\bibitem{SDN1}
S.~Sezer, S.~Scott-Hayward, P.~Chouhan, B.~Fraser, D.~Lake, J.~Finnegan,
  N.~Viljoen, M.~Miller, and N.~Rao, ``{Are we ready for SDN? Implementation
  challenges for software-defined networks},'' \emph{IEEE Commun. Mag.},
  vol.~51, no.~7, pp. 36--43, Jul. 2013.

\bibitem{CRAN3}
C.~Liu, K.~Sundaresan, M.~Jiang, S.~Rangarajan, and G.-K. Chang, ``{The Case
  for Re-configurable Backhaul in Cloud-RAN Based Small Cell Networks},'' in
  \emph{Proc. IEEE INFOCOM}, Apr. 2013, pp. 1124--1132.

\bibitem{femto2}
N.~Golrezaei, A.~Molisch, A.~Dimakis, and G.~Caire, ``{Femtocaching and
  device-to-device collaboration: A new architecture for wireless video
  distribution},'' \emph{IEEE Commun. Mag.}, vol.~51, no.~4, pp. 142--149,
  2013.

\bibitem{metics}
Metis, ``{Scenarios, requirements and KPIs for 5G mobile and wireless
  system},'' May 2013.

\bibitem{wang1}
J.~Wang, H.~Zhu, and N.~Gomes, ``{Distributed antenna systems for mobile
  communications in high speed trains},'' \emph{IEEE J. Sel. Areas Commun.},
  vol.~30, pp. 675--683, May 2012.

\bibitem{5G4}
Q.~Li, H.~Niu, A.~Papathanassiou, and G.~Wu, ``{5G Network Capacity: Key
  Elements and Technologies},'' \emph{IEEE Vehicular Technology Mag.}, vol.~9,
  no.~1, pp. 71--78, Mar. 2014.

\bibitem{ref2}
A.~Gelberger, N.~Yemini, and R.~Giladi, ``{Performance Analysis of
  Software-Defined Networking (SDN)},'' in \emph{IEEE 21st International
  Symposium on MASCOTS}, Aug. 2013.

\bibitem{Zipf}
M.~Cha, H.~Kwak, P.~Rodriguez, Y.~Ahn, and S.~Moon, ``{I Tube, You Tube,
  Everybody Tubes: Analyzing The World's Largest User Generated Content Video
  System},'' in \emph{Proc. ACM SIGCOMM Internet Measurement}, Oct. 2007.

\bibitem{kongtao}
K.~Wang, Z.~Chen, and H.~Liu, ``{Push-Based Wireless Converged Networks for
  Massive Multimedia Content Delivery},'' \emph{IEEE Trans. Wireless Commun.},
  vol.~13, no.~5, pp. 2894--2905, May 2014.

\bibitem{5G1}
X.~Wang, M.~Chen, T.~Taleb, A.~Ksentini, and V.~Leung, ``{Cache in The Air:
  Exploiting Content Caching and Delivery Techniques for 5G Systems},''
  \emph{IEEE Tans. Commun. Mag.}, vol.~52, no.~2, pp. 131--139, Feb. 2014.

\bibitem{5G3}
J.~Andrews, S.~Buzzi, W.~Choi, S.~Hanly, A.~Lozano, A.~Soong, and J.~Zhang,
  ``{What Will 5G Be?}'' \emph{IEEE JSAC}, vol.~PP, no.~99, Jun. 2014.

\bibitem{ICN1}
G.~Xylomenos, C.~Ververidis, V.~Siris, N.~Fotiou, C.~Tsilopoulos, X.~Vasilakos,
  K.~Katsaros, and G.~Polyzos, ``{A Survey of Information-Centric Networking
  Research},'' \emph{IEEE Commun. Surveys Tutorials}, vol.~16, no.~2, pp.
  1024--1049, July 2014.

\bibitem{scrability}
S.~Yeganeh, A.~Tootoonchian, and Y.~Ganjali, ``{On Scalability of
  Software-defined Networking},'' \emph{IEEE Commun. Mag.}, vol.~51, no.~2, pp.
  136--141, Feb. 2013.

\bibitem{twcmy}
C.~Yang, Y.~Yao, Z.~Chen, and B.~Xia, ``{Analysis on Cache-enabled Wireless
  Heterogeneous Networks},'' \emph{{[online] http://arxiv.org/abs/1508.02797}},
  Aug. 2015.

\end{thebibliography}
\end{document}